# Plasmonic properties of electrochromic doped metal oxides investigated through Kubelka-Munk formalism

Michaël Lobet[1,]*, Florian Gillissen[2,] *, Nicolas DeMoor[1,], Jennifer Dewalque[2], Pierre Colson[2], Rudi Cloots[2], Anthony Maho[2,3], Luc Henrard[1]

[1] Department of Physics and Namur Institute of Structured Materials, University of Namur, Rue de Bruxelles 61, 5000 Namur, Belgium
[2] Group of Research in Energy and Environment from Materials (GREEnMat), CESAM Research Unit, University of Liège, Allée du Six-Août 13, B-4000 Liège, Belgium.
[3] Univ. Bordeaux, CNRS, Bordeaux INP, ICMCB, UMR 5026, 87 Avenue du Docteur Schweitzer, 33600 Pessac, France.

*Those authors contributed equally
Corresponding author: michael.lobet@unamur.be

**Abstract**

Materials with broadband tunable optical properties are looked for in smart windows applications. Doped metal oxides presenting dual-band visible (VIS) – near-infrared (NIR) electrochromic properties can be used for solving such a challenge, and their accurate optical characterization is therefore of prime importance. Kubelka-Munk model is a state-of-the-art way to optically quantify the absorption properties of materials and is occasionally applied to plasmonic materials, even if great care should be taken to meet the formalism hypotheses. In the present work, Kubelka-Munk theory is discussed in the context of particles of indium-tin oxide and molybdenum-tungsten oxide formulations that are used as single-NIR and both-VIS/NIR active advanced electrochromic materials, respectively. An analytical model is derived for particles of much smaller dimensions than the incident wavelength and is experimentally verified. A dilution method is applied to verify the plasmonic characteristics of the particles. This study is key for efficient characterization of optical properties of metal oxides, and plasmonic materials in general, from diffuse reflectance measurements.

1. Introduction

Plasmonic materials have gained interest due to their enhanced light-matter interactions, impacting fields such as sensing, imaging, targeted drug delivery or photothermal therapy[1]. Among the variety of compositions of plasmonic particles, doped metal oxides have shown interesting properties for Surface-Enhanced Raman Scattering (SERS) spectroscopy, visible light enhanced catalytic reactions and electrochromic materials for smart windows applications. Smart windows based on plasmonic electrochromic compounds offer the possibility to fine tune the visible (VIS) and near-infrared (NIR) parts of the solar spectrum independently, which is not possible with conventional, non-plasmonic electrochromic materials[2]. Such dual-band operation is looked for increasing the energy savings in buildings, allowing one to manage selectively the luminosity and the heat via different modes of solar light modulation amongst bright (VIS- and NIR-transparent), cool (VIS-transparent, NIR-opaque), warm (VIS-opaque, NIR-transparent), or dark (VIS- and NIR-opaque) optical states[3].

Selected formulations of doped metal oxides such as indium-tin oxide ITO[4] or substoichiometric tungsten oxide $WO_{3-x}$[5,6] are particularly relevant in this context, respectively responding in the 1500-2000 nm and 800-1200 nm wavelength ranges. Their optical properties (absorption/transmission) need however to be finely characterized. A common experimental way to determine the absorbance spectrum of plasmonic particles is to disperse them in a suitable liquid solvent and measure absorption using Beer-Lambert law[7,8]. Though, such easy optical characterization fails in some case, for example when the processed liquid suspension is unstable.

The Kubelka-Munk model[9–11], originally derived for paintings[12], uses a simple diffuse reflectance to probe the absorption properties of solid-state samples. Due to its easy implementation for powders of particles, the model has been widely applied in scientific and industrial characterizations of materials such as paper[13,14], pigmented plastics[15], human tissues[16], teeth[17,18] or pigment mixtures including iron oxide[19]. It was also applied to a range of plasmonic materials[20–24]. However, one should not forget that the Kubelka-Munk formalism comes with specific hypotheses[11], such as a weakly absorbing medium necessary to minimize specular reflection. A careless use of the Kubelka-Munk formalism to highly absorbing medium, such as plasmonic particles around their localized surface plasmon resonance (LSPR), should be avoided. To the best of our knowledge, a detailed study on the conditions of application of Kubelka-Munk theory to plasmonic particles is missing from the current literature.

Therefore, in this work, we discuss the use of the Kubelka-Munk formalism and the validity of its hypotheses in a plasmonic context. We derive an analytical formula for computing the Kubelka-Munk function in the case of small plasmonic particles within the quasi-static approximation. This analytical derivation allows us to identify characteristic signatures in the Kubelka-Munk spectrum of plasmonic particles upon dilution in a suitable dielectric background. We produce different formulations of particles, i.e. indium-tin and molybdenum-tungsten oxides, and verify their plasmonic characteristics using diffuse reflectance spectrometry, according to the derived analytical model. These plasmonic particles are particularly interesting in the context of designing broadband optically-tunable materials, notably for electrochromic smart windows applications.

## 2. *Kubelka Munk model for plasmonic materials*

Optical properties of plasmonic particles processed as suspensions from colloidal synthesis approaches are classically determined through spectrophotometry. The obtained suspension is illuminated by an incident intensity $I_0$ while the emerging electromagnetic radiation of intensity $I < I_0$ is collected in a detector after going through an optical path $z$ in the particulate medium (Fig. 1a)[8].

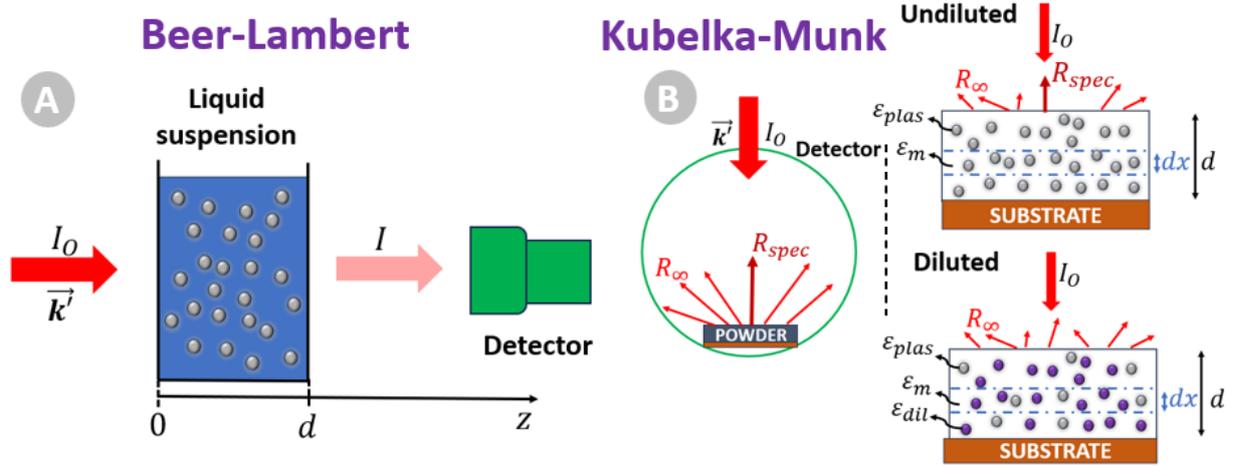

**Figure 1**: (a) Experimental setup for (a) Beer-Lambert absorption measurements in liquid suspension and (b) Kubelka Munk model for plasmonic particles (in grey), with both undiluted and diluted configurations (diluting agent in blue) inside an integrating sphere.

The experimental situation can be modeled as a collection of particles, more or less uniformly distributed within the region $0 \leq z \leq d$ with a non-absorbing background (in blue in Fig. 1a), and a number of particles per unit volume $N$. In this case, scattering is considered incoherent, the separations between the particles being uncorrelated during the measurement. Therefore, one can deduce the well-known Beer-Lambert law[8]:

$$I(z) = I_0 e^{-\mu_{ext} z} \qquad (1)$$

providing the intensity of attenuated light after a distance $z$ with the attenuation coefficient $\mu_{ext} = NC_{ext} = NC_{abs} + NC_{sca} = 2(k+s)$ where $C_{ext}$, $C_{abs}$ and $C_{sca}$ are the extinction, absorption and scattering cross-section and $k$ and $s$ the absorption and scattering coefficients, respectively. Using the Beer-Lambert law, one usually characterizes the optical properties of a (plasmonic) material by its absorbance $A = -log(T) = \frac{\mu_{ext} d}{ln10}$ with $T = \frac{I(d)}{I_0}$ being the transmittance through the suspension. This model considers a single unidirectional energy flux. Therefore, equation (1) and the deduction of the absorbance are valid for small contribution from the scattering to the total attenuation, *i.e.* $NC_{sca}d = sd \ll 1$. This condition is generally met in plasmonic materials when the size $r$ of the particles is smaller than the LSPR wavelength $\lambda = 2\pi/k'$ where $k'$ is the wave vector inside the medium, hence neglecting scattering in the quasi-static limit, *i.e.* $C_{ext} \approx C_{abs}$ if $k'r \ll 1$.

However, for suspensions unstable because of their composition and/or processing method, this methodology using dispersed particles in a solvent may fail at correctly providing optical characterization of plasmonic particles. Nevertheless, a relatively direct optical measurement can still provide a characterization of the absorption properties of a particulate medium if one measures the diffuse reflectance of the particles being processed as powders, potentially from the proper filtration and/or evaporation/drying of the aforementioned unstable suspensions, using the Kubelka-Munk (KM) formalism[11,25]. The later provides a direct link between the absorption coefficient $k$, scattering coefficient $s$ and the measured diffuse reflectance $R_\infty$ as:

$$KM = \frac{k}{s} = \frac{(1-R_\infty)^2}{2R_\infty}. \quad (2)$$

Using the above definitions for the absorption and scattering coefficients, the KM function is equal to $KM = \frac{C_{abs}}{C_{sca}}$. Provided that the scattering coefficient is constant[11], measuring the diffuse reflectance gives an easy way to access the absorption cross-section of a material. This is the reason why KM function is generally assimilated to (or confused with) an absorption spectrum.

Originally developed for characterizing paintings[12], special care should however be taken when used for characterizing plasmonic particles. A simple glance at equation (2) indicates that a vanishing scattering coefficient as for small particles – but not necessarily absorbing – would cause a divergence of the KM function, hence a wrong use of the equation (2) as is. A more detailed analysis requires recalling the underlying hypothesis of the KM model.

The KM model can be seen as an extension of the Beer-Lambert model because it considers energy flux in two opposite directions (Fig. 1b) rather than a unidirectional flux. We suppose a sample of powder of thickness $d$ divided into infinitesimal layers of thickness $dx$ relying on a substrate. The KM model is based on three hypotheses[11]:

(1) The particles inside the infinitesimal layer are considered as randomly distributed and with a size smaller than the thickness of the layer ($r \ll d$);
(2) The sample undergoes diffuse irradiation;
(3) The sample should display ideal diffuse reflection, *i.e.* reflection is isotropic, following Lambert cosine law[26,27]. Therefore, we can neglect specular reflection $R_{spec}$ in the total reflectance $R = R_{spec} + R_\infty \rightarrow R \sim R_\infty$.

The first hypothesis allows to consider the particle layer as homogeneous and avoids treating individual particles constituting the layer. Consequently, shapes and sizes of the particles can be non uniform[28]. The sufficiently thick layer also ensures that one can neglect the reflectance coming from the substrate. This assumption is usually met in realistic experimental conditions.

The second hypothesis enables the isotropic distribution of scattering within the layer[29]. In practice, no differences, within experimental errors, could be observed amongx three scenarios: (1) directive (45°) incident irradiation and directive vertical detection, (2) inside an integrating sphere, diffuse incident irradiation (*i.e.* the KM hypothesis) and vertical detection and (3) inside an integrating sphere, vertical incident irradiation and diffuse detection (our setup, as in Fig. 1b) [11].

In case of low reflection/high absorption, which is the case around plasmonic resonance, the applicability of the KM model is questionable since the third hypothesis is not met. Specular reflectance for an interface separating medium 1 with refractive index $n_1$ and an absorbing medium 2 with refractive index $\widetilde{n_2} = n_2 + i\kappa$ is easily obtained using Fresnel formula[30,31]. It yields for normal incident light:

$$R_{spec} = \frac{(n_2-n_1)^2 + \kappa^2}{(n_1+n_2)^2 + \kappa^2}. \quad (3)$$

At the limit of high $\kappa$, the specular reflectivity tends to one and $R_\infty$ is negligible. This remark is important for plasmonic systems which are highly absorbing at resonance. Therefore, directly applying the KM theory leads to unsuitable use of the model[11].

In order to counter this, using a dilution method enables to respect the KM hypothesis and properly characterize the absorption properties of the sample. It consists in grinding the absorbing powder and mixing it with a nonabsorbing standard such as LiF, NaCl, BaSO$_4$ or MgO, being added in excess so that it eliminates the original specular reflection. Consequently, the scattering coefficient of the mixture is almost solely determined by the scattering coefficient of the diluting agent [32]. Using the pure diluting agent as reference for the reflectance measurements enables to neglect residual specular reflection as well as possible deviations from the isotropic scattering distribution in such a way that the measurements become independent of the experimental setup. Kortum explicitly demonstrated that the KM spectrum from undiluted anthraquinone does not resemble the "real" KM spectrum obtained by diluting anthraquinone in ethanolic solution[32]. The real spectrum is only recovered once the powder is sufficiently diluted in NaCl. The same behavior occurs for plasmonic particles as shown below. Therefore, such dilution method is a convenient experimental tool to optically characterize absorbing powder and should be considered as a *sine qua non* condition for using the KM model with plasmonic particles.

Applying the dilution method to the KM model for plasmonic particles mixed with a diluent agent reaches

$$KM = \frac{k_{mix}}{s_{mix}} = \frac{k_{plas} + k_{dil}}{s_{plas} + s_{dil}} \cong \frac{k_{plas}}{s_{dil}} \quad (4)$$

where the indices $\square_{mix}, \square_{plas}, \square_{dil}$ stand for mixture, plasmonic and diluent agent respectively. We also assume that the diluent agent is non-absorbing ($k_{dil} \to 0$) and dominates the scattering coefficient of the mixture. The later assumption is further justified if the diluent is in coarse grain state (*i.e.* the size $r_{dil} > r_{plas} \to s_{dil} \gg s_{plas}$). Considering a constant scattering cross-section of the diluent agent $C_{sca,dil}$ [11], one obtains

$$KM \cong \frac{N_{plas}C_{abs,plas}}{N_{dil}C_{sca,dil}}. \quad (5)$$

Therefore, one can now correctly relate the diffuse reflectance measurement to the absorption cross-section of the plasmonic particles measured once diluted, meeting KM hypotheses. Crucially, a trade-off must be reached between a sufficient density of plasmonic particles $N_{plas}$ to have a measurable signal-to-noise ratio, and a not too high quantity to avoid specular reflection as discussed above.

Furthermore, one can provide an analytical formulation for the KM model for spherical plasmonic particles if one uses the electrostatic approximation [1,8], valid when the size (here the radius) of the plasmonic particles is small compared to the vacuum wavelength $r_{plas} \ll \lambda_0$:

$$KM \cong \frac{8\pi^2}{\lambda_0}\left(\frac{1}{C_{sca,dil}}\right)\frac{f_{plas}}{1-f_{plas}}r_{plas}^3\sqrt{\varepsilon_{m,eff}}Im\left(\frac{\varepsilon_{plas}-\varepsilon_{m,eff}}{\varepsilon_{plas}+2\varepsilon_{m,eff}}\right) \quad (6)$$

with $f_{plas}$ being the filling fraction of plasmonic particles, related to the mass percentage of plasmonic particles used in the dilution $m_{plas}(\%)$ and the mass densities $\rho$:

$$f_{plas} = \frac{m_{plas}(\%)}{m_{plas}(\%) + \left(\frac{\rho_{plas}}{\rho_{dil}}\right)\left(1 - m_{plas}(\%)\right)}. \tag{7}$$

The small spherical plasmonic particles have a permittivity $\varepsilon_{plas}$ embedded in an environment of permittivity $\varepsilon_m$ (Fig1b). The latter can be approximated by the effective permittivity $\varepsilon_{m,eff}$ using a Bruggeman model[33] to consider the effect of the varying quantities of plasmonic and diluent constituents upon dilution. The resonance in the analytical KM function corresponds solely to the Frölich condition $Re\left(\varepsilon_{plas}(\omega)\right) = -2\varepsilon_{m,eff}$, i.e. the dipole surface plasmon resonance of the sphere. It should be noted that changing the diluting agent has a direct influence on the plasmonic absorption peak: an increased surrounding permittivity induces a shift of the absorption maximum. More details and discussion on the derived model can be found in SM1.

### 3. Kubelka-Munk formalism applied to plasmonic indium-tin oxide

### 3.a Analytical model for plasmonic ITO

To further characterize the trends of the KM model for plasmonic particles upon dilution, let us apply the analytical model (equation (6)) to indium-tin oxide (ITO) as state-of-the-art formulation allowing for NIR-modulating plasmonic electrochromism [4,34–38]. We model here ITO nanoparticles of radius $r_{ITO} = 10nm$ and density $\rho_{ITO} = 7.18\,g/cm^3$ diluted in LiF with $\rho_{LiF} = 2.64\,g/cm^3, \varepsilon_{LiF} = 1.92$ [39]. We use a Drude model for the permittivity of ITO (see details in SM2 and Fig. S1). Here, $C_{sca,dil}$ is considered as a scaling factor as discussed in SM1. One can clearly see three major characteristics of the effect of increasing dilution of plasmonic particles, i.e. decreasing $m_{ITO}$ value, on the KM spectra (Fig. 2):
    (1) A red-shift of the peak of the KM function;
    (2) A diminution of the intensity of the KM function;
    (3) A broadening of the peak of the KM function.

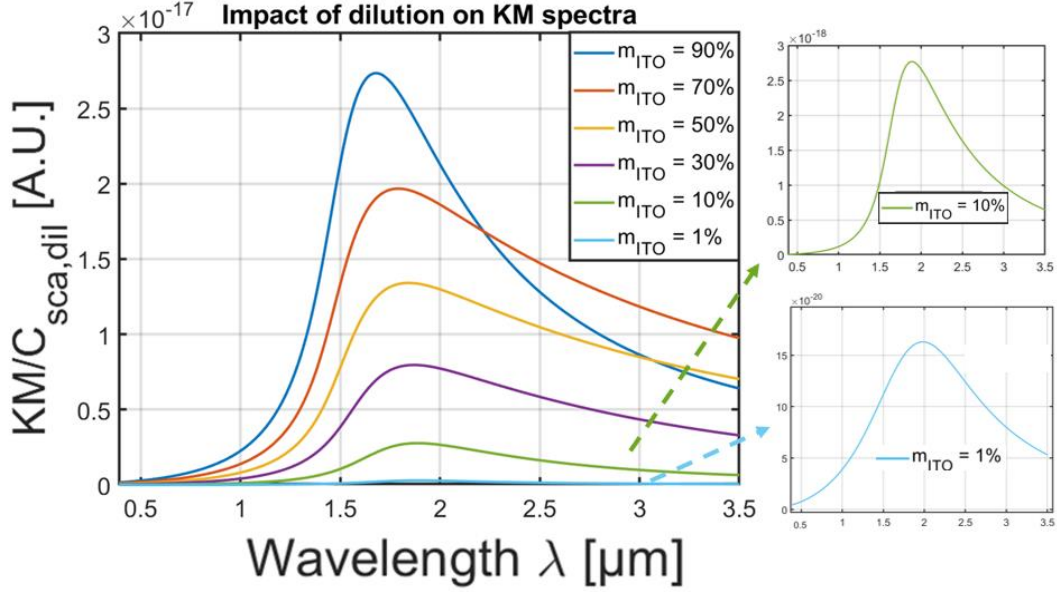

**Figure 2**: Impact of dilution of plasmonic particles, here ITO, on KM function. Insets: values for $m_{ITO} = 10\%$ and $m_{ITO} = 1\%$.

The derived analytical model for the KM function is dominated by the absorption cross section of the plasmonic particles as seen in equation (6), specifically with the terms $r_{plas}^3$, $\sqrt{\varepsilon_{m,eff}}$ and $Im\left(\frac{\varepsilon_{plas}-\varepsilon_{m,eff}}{\varepsilon_{plas}+2\varepsilon_{m,eff}}\right)$. The effective permittivity of the surrounding medium evolves from being dominated by ITO ($\varepsilon_{ITO} < 0$) to the one of LiF ($\varepsilon_{LiF} > 0$) upon dilution. This increased effective surrounding permittivity causes the red-shift as the frequency of the plasmonic peak for small spherical particles is given by $\omega_0 = \omega_p/\sqrt{1+2\varepsilon_m}$ with $\omega_p$ the plasma frequency[1]. This well-known result for absorption cross-section of plasmonic particles is illustrated in Fig. S2 (see SM2). The diminishing filling fraction of plasmonic particles once the mass percentage of those is lowered (equation (7)) upon dilution impacts the intensity. Lower number of plasmonic particles means lower $k_{plas}$ and higher $s_{dil}$ in equation (4), hence a globally diminishing intensity of the KM function. Finally, the broadening is related to the above mentioned Frölich condition for plasmonic resonance. If $\varepsilon_m$ increases, $Re(\varepsilon_{plas})$ becomes more negative, and correspondingly, in a Drude model, $Im(\varepsilon_{plas})$ increases (see Fig. S1, SM2), resulting in a broadening of the peak.

The impact of other parameters, such as modifications of the radius of the plasmonic particles $r_{plas}$, the plasma frequency $\omega_p$ or the damping constant $\gamma$, on the analytical model of KM for plasmonic particles is further discussed in SM2.

One can wonder for which filling fraction of plasmonic particles the KM hypotheses are specifically met. There is no universal answer to this question[11]. However, the present work can provide some guidelines. The main motivation of the dilution method was to meet the third hypothesis of the KM model, lowering the specular reflection and maximizing the diffuse reflectance. The quantity of plasmonic particles should be small enough to meet those criteria. It should be noted here that the second hypothesis of the KM model also supposes that particles are sufficiently separated from each other to neglect phase relations and interferences between the scattered waves[29]. Furthermore, we used a model for an isolated plasmonic sphere[1] considering the additive properties of the cross-sections, valid for incoherent scattering occurring once neglecting correlations between particles, *i.e.* at large interparticle separations. The above considerations lead to a quantity of plasmonic particles certainly below 50%. However, a too small percentage of plasmonic particles leads to a too low KM function because of the dominating $s_{dil}$ factor causing small signal to noise ratio in experiments. Therefore, a threshold should be found. At this stage, we need more experimental considerations to further answer this question and confront the derived analytical model to verification.

### *3.b. Experimental verification for plasmonic ITO*

To experimentally validate the proposed analytical KM model for plasmonic particles, ITO particles are fabricated through colloidal synthesis[38]. Well dispersed and defined nanospheres are obtained, with a narrow diameter distribution averaging 5.5 ± 1.0 nm, based on transmission electron microscopy (TEM) images recorded using a TECNAI G² 20 operated at 200 kV and shown on Fig. 3a. Particles in the form of powder are then obtained by heating the suspension at 100°C until a dry solid is left. The sample is manually crushed in a mortar and diluted with LiF powder (at various %wt.) until a homogeneous mix is obtained, which is then transferred into a suitable holder for further optical characterization using UV-VIS-NIR spectrometry. Any extra powder is removed so that the surface of the sample is leveled with the holder.

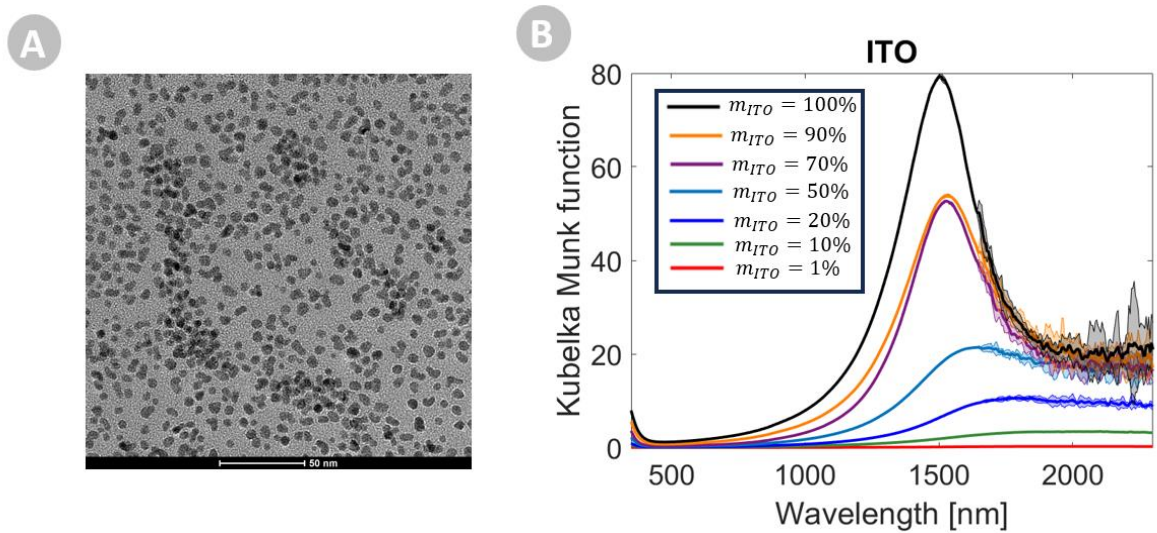

**Figure 3**: (a) TEM image of ITO particles, (b) KM function of ITO particles using LiF as diluting agent.

Diffuse reflective measurements by UV−VIS−NIR spectrometry are performed on a Shimadzu 3600 Plus instrument using an integrating sphere (ISR-1503). Data is then post-processed using equation (2) to obtain experimental KM function. We performed 5 measurements, their mean value as well as the 99.75% confidence level are shown on Fig. 3b. The curve for $m_{ITO} = 100\%$ corresponds to undiluted ITO but does not fit the above explained KM hypotheses, so it should be considered carefully.

Upon dilution in LiF, the three main characteristics of KM function presented in the proposed model are recovered, namely (1) the red-shift of the peak, (2) the diminution of the intensity and (3) the broadening of the peak of the KM function for a progressively lowered mass of plasmonic particles. Peak and FWHM values are reported in SM3 (Table 1). The peak shifts from 1495 nm ($m_{ITO} = 90\%$) to 2040 nm ($m_{ITO} = 10\%$).

It could be wrongly deduced, by a simple glance at Fig. 3, that the noise attenuates with a lower quantity of ITO. This is not true, as it results from the attenuation of the KM signal (2). A normalized version of Fig. 3 is presented in SM 4 (Fig. S6). The noise above 1700 nm is due to the PbS NIR detector which is sensitive to any hot objects in the environment. Moreover, the diffuse reflectivity of the mixed powder is very small in that wavelength range for $m_{ITO} > 20\%$ (see SM4 Fig. S7), decreasing the signal-to-noise ratio.

In the following sections, we will consider $m = 10\%$ as a relevant threshold to verify the KM hypotheses and having sufficient signal to detect: the diffuse reflectance curve for $m = 10\%$ is well separated from the others, higher than the noise and with an intensity not close from zero.

All in all, the above results experimentally validate qualitatively the proposed analytical model and provide guidelines to properly apply the KM model to other plasmonic particles.

## 4. *Application to molybdenum-tungsten oxides as emergent electrochromic materials*

Interestingly, recent works have highlighted the occurrence of a largely-increased plasmonic absorption in other formulations of doped metal oxides, notably in oxygen-deficient molybdenum – tungsten mixed oxides being praised in photocatalytic applications. Specifically, Yin *et al.* have shown that the KM function of these $Mo_{1-y}W_yO_{3-\delta}$ – "MoWOx" – hybrids, presented KM intensities being 20 and 16 times larger than KM of parent formulations $MoO_{3-x}$ and $WO_{3-x}$, respectively[24]. They claimed that the enhanced absorption is a LSPR signal arising from vacancies in the crystal structure. They mainly studied the molar ratio of Mo:W = 1:1, but the 2:1 and 1:2 ratios presented red-shifted KM peaks as well as decreased peak intensities of about 20% and 50% vs. 1:1, respectively[24].

These "MoWOx" formulations further present noticeable interest as advanced electrochromic compounds (on-going studies), as they show enhanced electrochromic efficiency and dual-band selectivity towards both VIS and NIR wavelength ranges, highlighting the processed formulations as promising candidates for the development of "new generation" dual – band smart windows.

Therefore, we propose here to apply the KM formalism on these compounds to check the presence of plasmonic character.

We prepared MoWOx hybrids as well as parent oxides $MoO_{3-x}$ and $WO_{3-x}$, adapting synthesis protocols from literature[24]. Briefly, the materials are obtained through a single step solvothermal synthesis: 9 mmol of metallic powder precursors are dissolved in 11.5 mL of $H_2O_2$, then mixed with 69 mL of isopropanol and transferred into a 125 mL Teflon body, placed into a stainless-steel autoclave to be thermally treated for 1h at 160°C. For the MoWOx, 6 mmol of Mo and 3 mmol of W precursor are used (Mo/W : 2/1), while $WO_{3-x}$ and $MoO_{3-x}$ parent oxides are produced from 9 mmol of W and 9 mmol of Mo metallic powder, respectively. The obtained MoWOx particles present a peculiar "urchin-like" morphology (see TEM image in Fig. 4a), consisting in a solid core, with its surface covered by nanorods. These particles show average dimensions of 1.5 ± 0.5 µm. Parent oxides $MoO_{3-x}$ and $WO_{3-x}$ exhibit different morphologies, with $MoO_{3-x}$ (Fig. 4b) presenting a feather-like morphology (elongated platelets of dimensions 1.9 x 0.1 x 0.05 ± 0.5 x 0.03 x 0.01µm) while its W counterpart is found as aggregated nanospheres with an average diameter of 6.4 ± 1.6 nm (Fig. 4c). It is interesting to note here that, although the shape, size and composition of the particles may be different, general conclusions and trends arising from the analytical model derived in section 2 are still valid, because those parameters are contained in equation 6, thus impacting the KM function.

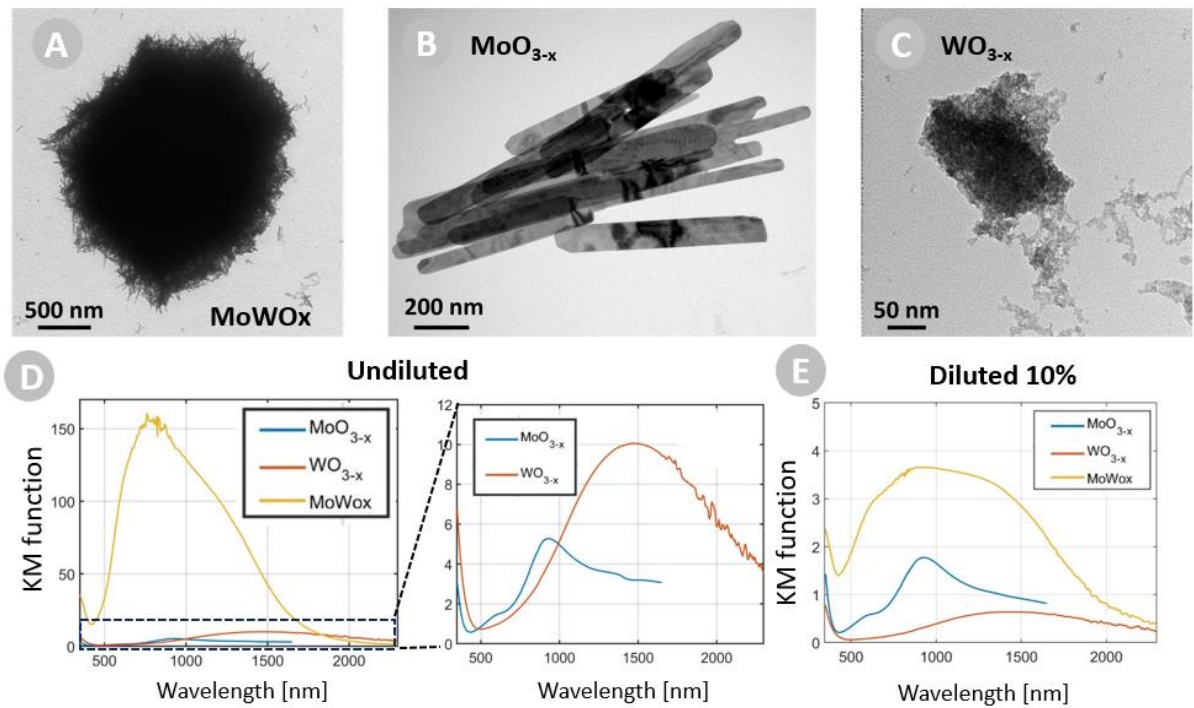

**Figure 4**: TEM images of (a) MoWOx, (b) $MoO_{3-x}$ and (c) $WO_{3-x}$ particles. (d) KM function of undiluted MoWox particles and parent oxides $MoO_{3-x}$ and $WO_{3-x}$. Inset: Zoom on KM function for parent oxides (e) KM function for MoWOx particles and parent oxides $MoO_{3-x}$ and $WO_{3-x}$ with 10% of mass percentage for the particles of interest.

| Studied particle | Dilution | Peak intensity [a.u.] | Peak wavelength [nm] |
|---|---|---|---|
| MoWOx | Undiluted *This study* | 160 | 760 |
| | Undiluted *Yin[24]* | 130 | 655 |
| | Diluted 10% *This study* | 3.6 | 950 |
| $MoO_{3-x}$ | Undiluted *This study* | 5 | 929 |
| | Undiluted *Yin[24]* | 8 | 920 |
| | Diluted 10% *This study* | 1.8 | 927 |
| $WO_{3-x}$ | Undiluted *This study* | 10 | 1485 |
| | Undiluted *Yin[24]* | >10 | >1400 |
| | Diluted 10% *This study* | 0.6 | 1450 |

**Table 1**: Comparison of KM peak intensities and peak wavelengths of MoWOx, $MoO_{3-x}$ and $WO_{3-x}$ particles studied in this work, compared to measurement from Yin *et al*[24]. We also added diluted measurements with $m_{plas} = 10\%$.

First, let us compare, as Yin *et al.* did[24], the undiluted KM function of MoWOx hybrid and parent oxides $MoO_{3-x}$ and $WO_{3-x}$ (Fig. 4d and Table 1). The peak intensities are of 160 a.u. at 760 nm, 5 a.u. at 929 nm and 10 a.u. at 1485 nm, respectively. As comparison, Yin *et al.* reported KM peak intensities around 130 a.u. at 655 nm for MoWOx 2:1, 8 a.u. at 920 nm for $MoO_{3-x}$ and above 10 a.u. and above 1400 nm (out of measurement range) for $WO_{3-x}$. Our results are therefore in good agreement with literature ones.

However, as explained above, a better comparison would be for a dilution with a mass of the studied species respecting the previously defined threshold, here of 10% compared to the LiF diluting agent, as reproduced on Fig. 4e. $MoO_{3-x}$ peaks at 927 nm while $WO_{3-x}$ peaks at 1450 nm. The KM function of MoWOx peaks at 950 nm with a doubled intensity compared to $MoO_{3-x}$, and six times increase compared to $WO_{3-x}$. The enhancement of absorption is clearly less spectacular than with undiluted samples but does correspond to fulfilled KM hypotheses. The broadband character of MoWOx is clearly evidenced, with enhanced contributions in both VIS and NIR ranges compared to parent oxides. No significant peak shifts are identified for parent oxides compared to undiluted species, while MoWOx appears red-shifted. This suggests a plasmonic behavior in NIR for MoWOx while low- or non-plasmonic character for parent oxides.

To verify this plasmonic property of MoWOx metal oxides, let us check if the KM spectra verify the three major criteria of the analytical model as presented in section 3. While the diminution of the KM signal upon dilution, *i.e.* criteria (2), is apparent on Fig 5a, the red-shift of the peak as well as its broadening are not obvious here. A normalized KM function is presented on Fig. 5b to better assess those characteristics.

The broadening (480 nm of FWHM) and the red-shift (+155 nm of LSPR maximum) are now clearly visible as reported data in SM5 (Table 2). We may further infer from those data that the dual-band electrochromic property arises from a combined effect of polaronic and plasmonic mechanisms[40]. To further illustrate that hypothesis, we realized a fit of the experimental data for both $m_{MoWOx} = 100\%$ (undiluted) and $m_{MoWOx} = 10\%$ with a deconvolution of two Gaussian curves (Fig. 5c). The first Gaussian fits peak at respectively 735 nm for undiluted and 745 nm for $m_{MoWOx} = 10\%$, while the second ones peak at 1055 nm (undiluted) and 1350 nm (10%). It is consistent with an interpretation of a polaritonic behavior resulting in non-shifted peak contributions (first gaussian at lower wavelength) and a plasmonic behavior resulting in red-shifted peak contributions (second gaussian at higher wavelength).

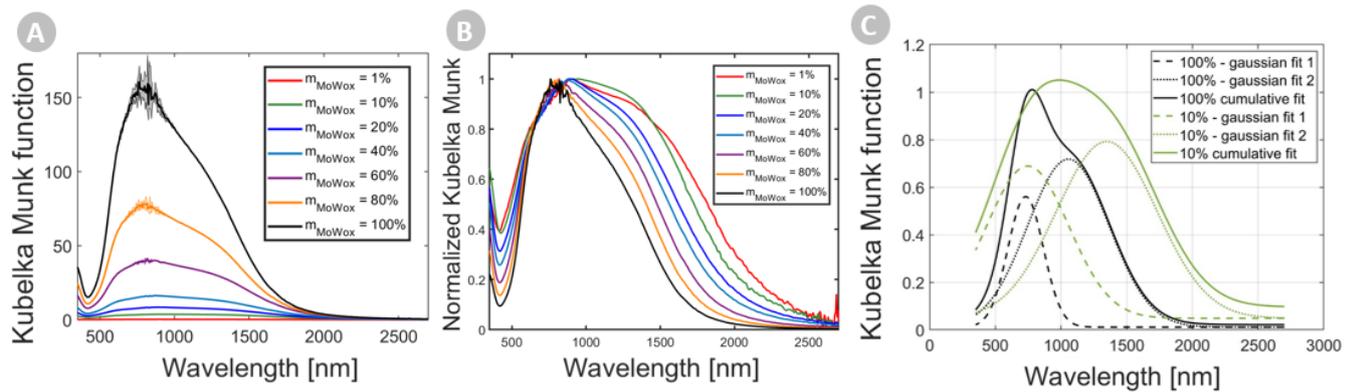

**Figure 5**: (a) KM function of MoWOx particles upon dilution. (b) Normalized KM function of MoWOx particles upon dilution. (c) KM function for undiluted MoWOx ($m_{MoWOx} = 100\%$, black curves) and diluted at 10% (green curves), with gaussian fits separating the global KM spectra in 2 components (gaussian fits 1 and 2 for each).

To further discuss the occurrence of plasmonic behavior in metal oxide particles using the presented KM formalism, one can study the modification of the KM spectra with a different diluting agent. As expressed by our analytical formulation (equation 6) and shown in SM1, an increased electric permittivity background would cause a red-shift of the KM spectra. To illustrate this, we use BaSO$_4$ whose permittivity is $\varepsilon_{BaSO_4} = 265 > \varepsilon_{LiF} = 1.924$[37]. Here again, we check this affirmation with state-of-the-art ITO particles first (Fig. 6a) and indeed KM function for BaSO$_4$ appears red-shifted compared to LiF diluting agent, itself being red-shifted compared to undiluted KM spectra. Similar trends occur for MoWOx, further confirming its plasmonic behavior (Fig. 6b).

As the optical properties of MoWOx particles is supposed to result from oxygen vacancies, one might finally anneal the powder to suppress those vacancies. As seen on Fig. 6c, "stoichiometric MoWOx" *i.e.* Mo$_{1-y}$W$_y$O$_3$ or "(Mo,W)O$_3$", resulting from substoichiometric MoWOx compounds being thermally annealed at 500°C for 12h under air (so to suppress the oxygen vacancies), has a severe drop in KM intensity This much less intense KM intensity of "stoichiometric MoWOx" shows no significant trends with the dilution in LiF and BaSO4 (Fig 6d). Fully oxidized "stoichiometric MoWOx" can then be considered as non-plasmonic.

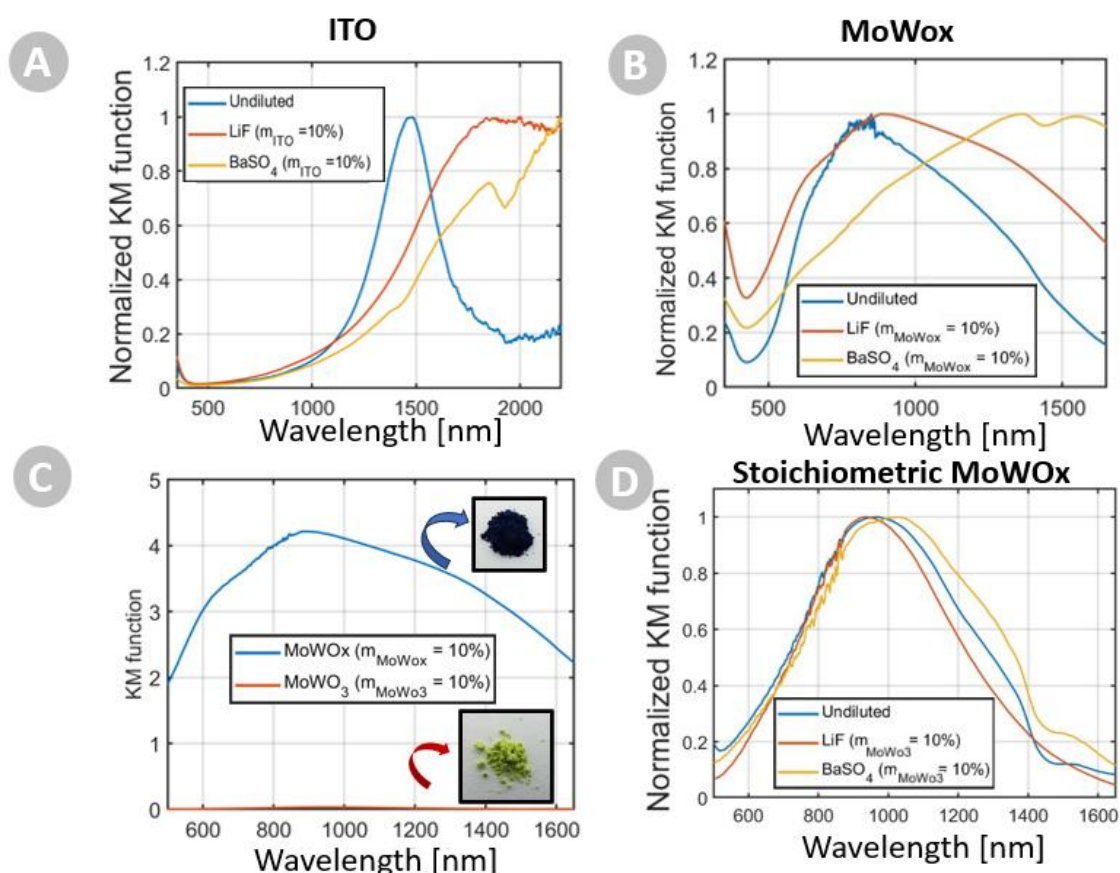

**Figure 6**: KM function for undiluted and different diluting agents (LiF and BaSO$_4$) for (a) ITO and (b) MoWOx particles. (c) Comparison of KM spectra of MoWOx and MoWO$_3$ or "stoichiometric MoWOx". Insets: optical photographs of MoWOx (up) and (Mo,W)O$_3$ (down). (d) KM function for (Mo,W)O$_3$ being undiluted and using different diluting agents (LiF and BaSO$_4$).

## 6. Conclusions

In conclusion, we thoroughly investigated the application of a state-of-the-art model, the Kubelka-Munk formalism, to a specific problem of topical context, the optical characterization of plasmonic particles used for electrochromic applications. We highlighted the importance of using the dilution method to respect the original Kubelka-Munk hypotheses. An analytical model of KM ratio for plasmonic particles is provided within the limit of small particles (quasi-static approximation). This model allows to identify specific characteristics of the evolution of the KM function upon dilution, such as a red-shift and a broadening of the peak of the KM function as well as a diminution of the global absorption intensity of the KM function. Moreover, the KM function is also shown to red-shift when the diluting agent is changed to a higher dielectric environment. All those characteristics have been verified both numerically and experimentally on plasmonic ITO particles to assess the validity of the proposed analytical model. Moreover, it was applied to emergent doped metal oxides such as MoWOx hybrids. Those MoWOx hybrids were compared to parent $MoO_{3-x}$ and $WO_{3-x}$ formulations, leading to ameliorated practice, and understanding in approaches of optical characterization of (non-)plasmonic particles. Those findings are notably useful in the context of electrochromic materials and could be applied to other electrochromic metal oxide formulation such as NiO or other shapes and structures. The present results are general enough to be applied to other plasmonic particles, beyond the present context of electrochromism.

**Authors contribution**

ML derived the analytical modelling helped by NDM and LH. FG carried out the sample fabrications under guidance and supervision of AM, JD, PC and RC, and optical measurements with advice from ML and NDM. NDM performed the Drude model analysis and the preliminary discussions on KM model. ML, FG, NDM, AM and LH analyzed all the data. LH, RC, ML and AM conceived the idea, suggested the designs, planned, coordinated and directly supervised the work. ML wrote the first draft of the manuscript, FG, AM and LH edited it and all authors contributed to the writing of the final version.

**Acknowledgments**
The authors acknowledge the support of the FRS-F.N.R.S. under the convention PDR T.0125.30 PLASMON_EC. M.L. is a Research Associate of the Fonds de la Recherche Scientifique – FNRS. This research used resources of the "Plateforme Technologique de Calcul Intensif (PTCI)" (http://www.ptci.unamur.be ) located at the University of Namur, Belgium, which is supported by the FNRS-FRFC, the Walloon Region, and the University of Namur (Conventions No. 2.5020.11, GEQ U.G006.15, 1610468, RW/GEQ2016 et U.G011.22). The PTCI is member of the "Consortium des Équipements de Calcul Intensif (CÉCI)" (http://www.ceci-hpc.be). A.M. acknowledges University of Bordeaux and ICMCB, where he holds a Junior Professor Chair, for additional support and funding. The authors would like to acknowledge the discussions with Antoine Honet and Richard Ngamkam as well as fruitful interactions and longstanding collaborations with Michel Voué. Moreover, the authors warmly thank Jean-François Colomer for TEM images.

**References**

(1) Maier, S. A. *Plasmonics: Fundamentals and Applications*; Springer US: New York, NY, 2007. https://doi.org/10.1007/0-387-37825-1.
(2) Granqvist, C. G.; Lansåker, P. C.; Mlyuka, N. R.; Niklasson, G. A.; Avendaño, E. Progress in Chromogenics: New Results for Electrochromic and Thermochromic Materials and Devices. *Sol. Energy Mater. Sol. Cells* **2009**, *93* (12), 2032–2039. https://doi.org/10.1016/j.solmat.2009.02.026.
(3) Bai, T.; Li, W.; Fu, G.; Zhang, Q.; Zhou, K.; Wang, H. Dual-Band Electrochromic Smart Windows towards Building Energy Conservation. *Sol. Energy Mater. Sol. Cells* **2023**, *256*, 112320. https://doi.org/10.1016/j.solmat.2023.112320.

(4) Garcia, G.; Buonsanti, R.; Runnerstrom, E. L.; Mendelsberg, R. J.; Llordes, A.; Anders, A.; Richardson, T. J.; Milliron, D. J. Dynamically Modulating the Surface Plasmon Resonance of Doped Semiconductor Nanocrystals. *Nano Lett.* **2011**, *11* (10), 4415–4420. https://doi.org/10.1021/nl202597n.

(5) Yan, J.; Wang, T.; Wu, G.; Dai, W.; Guan, N.; Li, L.; Gong, J. Tungsten Oxide Single Crystal Nanosheets for Enhanced Multichannel Solar Light Harvesting. *Adv. Mater.* **2015**, *27* (9), 1580–1586. https://doi.org/10.1002/adma.201404792.

(6) Zhang, S.; Cao, S.; Zhang, T.; Yao, Q.; Fisher, A.; Lee, J. Y. Monoclinic Oxygen-Deficient Tungsten Oxide Nanowires for Dynamic and Independent Control of near-Infrared and Visible Light Transmittance. *Mater. Horiz.* **2018**, *5* (2), 291–297. https://doi.org/10.1039/C7MH01128H.

(7) Mayerhöfer, T. G.; Popp, J. Beyond Beer's Law: Revisiting the Lorentz-Lorenz Equation. *ChemPhysChem* **2020**, *21* (12), 1218–1223. https://doi.org/10.1002/cphc.202000301.

(8) Bohren, C. F.; Huffman, D. R. *Absorption and Scattering of Light by Small Particles*, 1st ed.; Wiley, 1998. https://doi.org/10.1002/9783527618156.

(9) Yang, L.; Kruse, B. Revised Kubelka–Munk Theory I Theory and Application. *J. Opt. Soc. Am. A* **2004**, *21* (10), 1933. https://doi.org/10.1364/JOSAA.21.001933.

(10) Myrick, M.; Simcock, M.; Baranowski, M.; Brooke, H.; Morgan, S.; McCutcheon, J. The Kubelka-Munk Diffuse Reflectance Formula Revisited. *Appl. Spectrosc. Rev.* **2011**, *46* (2), 140–165. https://doi.org/10.1080/05704928.2010.537004.

(11) Kortüm, G. *Reflectance Spectroscopy*; Springer: Berlin, Heidelberg, 1969. https://doi.org/10.1007/978-3-642-88071-1.

(12) P, K. Ein Beitrag Zur Optik Der Farbanstriche. *Z Tech Phys* **1931**, *12*, 593–601.

(13) Džimbeg-Malčić, V.; Barbarić-Mikočević, Ž.; Itrić, K. Kubelka-Munk Theory in Describing Optical Properties of Paper ( I ). *Teh. Vjesn.* **2011**, *18* (1), 117–124.

(14) Dzimbeg-Malcic, V.; Barbarić-Mikočević, Ž.; Itrić, K. KUBELKA-MUNK THEORY IN DESCRIBING OPTICAL PROPERTIES OF PAPER (II). *Teh. Vjesn.-Tech. Gaz.* **2012**.

(15) Saunderson, J. L. Calculation of the Color of Pigmented Plastics*. *J. Opt. Soc. Am.* **1942**, *32* (12), 727. https://doi.org/10.1364/JOSA.32.000727.

(16) Van Gemert, M. J. C.; Welch, A. J.; Star, W. M.; Motamedi, M.; Cheong, W. F. Tissue Optics for a Slab Geometry in the Diffusion Approximation. *Lasers Med. Sci.* **1987**, *2* (4), 295–302. https://doi.org/10.1007/BF02594174.

(17) Ragain, J. C.; Johnston, W. M. Accuracy of Kubelka-Munk Reflectance Theory Applied to Human Dentin and Enamel. *J. Dent. Res.* **2001**, *80* (2), 449–452. https://doi.org/10.1177/00220345010800020901.

(18) Mikhail, S. S.; Azer, S. S.; Johnston, W. M. Accuracy of Kubelka–Munk Reflectance Theory for Dental Resin Composite Material. *Dent. Mater.* **2012**, *28* (7), 729–735. https://doi.org/10.1016/j.dental.2012.03.006.

(19) Barron, V.; Torrent, J. Use of the Kubelka—Munk Theory to Study the Influence of Iron Oxides on Soil Colour. *J. Soil Sci.* **1986**, *37* (4), 499–510. https://doi.org/10.1111/j.1365-2389.1986.tb00382.x.

(20) Pathak, T. K.; Swart, H. C.; Kroon, R. E. Structural and Plasmonic Properties of Noble Metal Doped ZnO Nanomaterials. *Phys. B Condens. Matter* **2018**, *535*, 114–118. https://doi.org/10.1016/j.physb.2017.06.074.

(21) Tanaka, A.; Hashimoto, K.; Kominami, H. Preparation of Au/CeO$_2$ Exhibiting Strong Surface Plasmon Resonance Effective for Selective or Chemoselective

Oxidation of Alcohols to Aldehydes or Ketones in Aqueous Suspensions under Irradiation by Green Light. *J. Am. Chem. Soc.* **2012**, *134* (35), 14526–14533. https://doi.org/10.1021/ja305225s.

(22) Kamimura, S.; Yamashita, S.; Abe, S.; Tsubota, T.; Ohno, T. Effect of Core@shell (Au@Ag) Nanostructure on Surface Plasmon-Induced Photocatalytic Activity under Visible Light Irradiation. *Appl. Catal. B Environ.* **2017**, *211*, 11–17. https://doi.org/10.1016/j.apcatb.2017.04.028.

(23) Wang, Z.; Liu, J.; Chen, W. Plasmonic Ag/AgBr Nanohybrid: Synergistic Effect of SPR with Photographic Sensitivity for Enhanced Photocatalytic Activity and Stability. *Dalton Trans.* **2012**, *41* (16), 4866–4870. https://doi.org/10.1039/C2DT12089E.

(24) Yin, H.; Kuwahara, Y.; Mori, K.; Cheng, H.; Wen, M.; Huo, Y.; Yamashita, H. Localized Surface Plasmon Resonances in Plasmonic Molybdenum Tungsten Oxide Hybrid for Visible-Light-Enhanced Catalytic Reaction. *J. Phys. Chem. C* **2017**, *121* (42), 23531–23540. https://doi.org/10.1021/acs.jpcc.7b08403.

(25) Bourdin, M.; Mjejri, I.; Rougier, A.; Labrugère, C.; Cardinal, T.; Messaddeq, Y.; Gaudon, M. Nano-Particles (NPs) of WO3-Type Compounds by Polyol Route with Enhanced Electrochromic Properties. *J. Alloys Compd.* **2020**, *823*, 153690. https://doi.org/10.1016/j.jallcom.2020.153690.

(26) Lambert, J.-H. *Photometria sive de mensura et gradibus luminis, colorum et umbrae*; Sumptibus Viduae Eberhardi Klett, typis Christophori Petri Detleffsen, 1760.

(27) Pedrotti, F. L.; Pedrotti, L. S. *Introduction to Optics*; Englewood Cliffs, N.J. : Prentice Hall, 1993.

(28) Danckwerts, P. V. Angewandte Chemie. *Chem. Eng. Sci.* **1962**, *17* (11), 955. https://doi.org/10.1016/0009-2509(62)87032-8.

(29) Theissing, H. H. Macrodistribution of Light Scattered by Dispersions of Spherical Dielectric Particles*. *J. Opt. Soc. Am.* **1950**, *40* (4), 232. https://doi.org/10.1364/JOSA.40.000232.

(30) Stenzel, O. *The Physics of Thin Film Optical Spectra: An Introduction*; Springer Series in Surface Sciences; Springer International Publishing: Cham, 2016; Vol. 44. https://doi.org/10.1007/978-3-319-21602-7.

(31) John David Jackson. *Classical Electrodynamics, 3nd Edition*; 1999.

(32) Kortüm, G.; Oelkrug, D. Über Den Streukoeffizienten Der Kubelka-Munk-Theorie. *Z. Für Naturforschung A* **1964**, *19* (1), 28–37. https://doi.org/10.1515/zna-1964-0107.

(33) Sihvola, A. H. *Electromagnetic Mixing Formulas and Applications*, Repr. with new cover.; IEE electromagnetic waves series; Institution of Electrical Engineers: London, 2008.

(34) Mendelsberg, R. J.; Garcia, G.; Milliron, D. J. Extracting Reliable Electronic Properties from Transmission Spectra of Indium Tin Oxide Thin Films and Nanocrystal Films by Careful Application of the Drude Theory. *J. Appl. Phys.* **2012**, *111* (6), 063515. https://doi.org/10.1063/1.3695996.

(35) Lounis, S. D.; Runnerstrom, E. L.; Bergerud, A.; Nordlund, D.; Milliron, D. J. Influence of Dopant Distribution on the Plasmonic Properties of Indium Tin Oxide Nanocrystals. *J. Am. Chem. Soc.* **2014**, *136* (19), 7110–7116. https://doi.org/10.1021/ja502541z.

(36) Mashkov, O.; Körfer, J.; Eigen, A.; Yousefi-Amin, A.-A.; Killilea, N.; Barabash, A.; Sytnyk, M.; Khansur, N.; Halik, M.; Webber, K. G.; Heiss, W. Effect of Ligand Treatment on the Tuning of Infrared Plasmonic Indium Tin Oxide

Nanocrystal Electrochromic Devices. *Adv. Eng. Mater.* **2020**, *22* (9), 2000112. https://doi.org/10.1002/adem.202000112.

(37) Maho, A.; Comeron Lamela, L.; Henrist, C.; Henrard, L.; Tizei, L. H. G.; Kociak, M.; Stéphan, O.; Heo, S.; Milliron, D. J.; Vertruyen, B.; Cloots, R. Solvothermally-Synthesized Tin-Doped Indium Oxide Plasmonic Nanocrystals Spray-Deposited onto Glass as near-Infrared Electrochromic Films. *Sol. Energy Mater. Sol. Cells* **2019**, *200*, 110014. https://doi.org/10.1016/j.solmat.2019.110014.

(38) Maho, A.; Saez Cabezas, C. A.; Meyertons, K. A.; Reimnitz, L. C.; Sahu, S.; Helms, B. A.; Milliron, D. J. Aqueous Processing and Spray Deposition of Polymer-Wrapped Tin-Doped Indium Oxide Nanocrystals as Electrochromic Thin Films. *Chem. Mater.* **2020**, *32* (19), 8401–8411. https://doi.org/10.1021/acs.chemmater.0c02399.

(39) Li, H. H. Refractive Index of Alkali Halides and Its Wavelength and Temperature Derivatives. *J. Phys. Chem. Ref. Data* **1976**, *5* (2), 329–528. https://doi.org/10.1063/1.555536.

(40) Tandon, B.; Lu, H.-C.; Milliron, D. J. Dual-Band Electrochromism: Plasmonic and Polaronic Mechanisms. *J. Phys. Chem. C* **2022**, *126* (22), 9228–9238. https://doi.org/10.1021/acs.jpcc.2c02155.